# Offloading Optimization with Delay Distribution in the 3-tier Federated Cloud, Edge, and Fog Systems

Ren-Hung Hwang[1], Yuan-Cheng Lai[2] and Ying-Dar Lin[3]
[1]Department of Computer Science and Information Engineering, National Chung Cheng University, Taiwan
[2]Department of Information Management, National Taiwan University of Science and Technology, Taiwan
[3]Department of Computer Science, National Chiao Tung University, Taiwan

rhhwang@cs.ccu.edu.tw, laiyc@cs.ntust.edu.tw, ydlin@cs.nctu.edu.tw

*Abstract*—Mobile edge computing and fog computing are promising techniques providing computation service closer to users to achieve lower latency. In this work, we study the optimal offloading strategy in the three-tier federated computation offloading system. We first present queueing models and closed-form solutions for computing the service delay distribution and the probability of the delay of a task exceeding a given threshold. We then propose an optimal offloading probability algorithm based on the sub-gradient method. Our numerical results show that our simulation results match very well with that of our closed-form solutions, and our sub-gradient-based search algorithm can find the optimal offloading probabilities. Specifically, for the given system parameters, our algorithm yields the optimal QoS violating probability of 0.188 with offloading probabilities of 0.675 and 0.37 from Fog to edge and from edge to cloud, respectively.

## I. Introduction

As mobile devices and mobile services increase significantly, 5G networks have been proposed to raise the transmission rate, increase the number of connections, and reduce transmission latency. 5G networks provide three application scenarios, enhanced Mobile Broadband (eMBB), Ultra-Reliable and Low Latency Communications (URLLC), and massive Machine Type Communications (mMTC). URLLC is a scenario that requests demand an ultra-low latency, such as autonomous vehicles and telemedicine.

Traditional Mobile Cloud Computing (MCC), which is two-tiered and consists of a cloud server and User Equipment (UEs), cannot provide ultra-low latency because the distance between the cloud server and UEs is usually over 100 kilometers. Instead of a cloud server, the edge servers of mobile networks provide the services and computing capacity by Mobile Edge Computing (MEC) techniques [1]. The main features of MEC are to push mobile computing, network control, and storage to the network edges. With MEC, the congestion in the core network can be alleviated, and the transmission delay can be reduced, significantly decreasing the total latency on obtaining services. Thus, a three-tier architecture with a federated cloud server, edge servers, and UEs appears.

In a multi-tier architecture, offloading to where, i.e., cloud server, edge server, or UE itself, to execute a request is an important issue. There were many previous studies focusing on the offloading decision problem in a two-tier architecture [2-9] or in a three-tier architecture [10-13]. These studies adopted offloading to reduce the latency or relieve the UE's energy consumption.

However, all previous papers which focused on latency reduction only considered the latency of each request or the average latency of all requests. However, for each request, it cares about whether its Quality of Services (QoS) requirement can be satisfied or not; namely, its latency can be less than a given threshold that the service needs. Thus, even the offloading algorithms or mechanisms of these studies actually reduce the latency; they still cannot reduce the probability of QoS violation, i.e., the latency exceeding a given threshold.

This paper considered the probabilistic offloading strategy in a three-tier federated architecture with the cloud server, edge servers, and UEs. When a computation request is generated at a UE, the probabilistic offloading strategy offloads the computation task to the edge server with probability $p^{UE}$ or performs the task at the UE with probability ($1-p^{UE}$). Similarly, when a computation request arrives at the edge server, the offloading strategy offloads the task to the cloud server with probability $p^{EC}$. The QoS requirement of computation tasks is its service delay, include communication delay and computation delay, which needs to be less than a given threshold. Our problem is to determine the optimal value of $p^{UE}$ and $p^{EC}$ such that the probability of violating the delay constraint of a task can be minimized. Queueing models and optimal algorithm for searching $p^{UE}$ and $p^{EC}$ are proposed to solve the problem. The main contributions of this study are:

(1) Closed-form solutions for calculating the probability of QoS violation are derived based on queueing models. We believe that this is the first work that formulates the federated three-tier computation offloading system using queueing models and derives a closed-form solution for the probability of violating the delay constraint.

(2) A novel sub-gradient search algorithm is proposed to find the optimal offloading probability.

The remainder of this paper is organized as follows: Section II describes the system architecture and system model. Closed-form solutions for calculating the probability of QoS violation are derived based on queueing models. The research problem is then defined; Section III presents the proposed Sub-Gradient Search (SGS) algorithm for finding the optimal solution.

Numerical results are then presented in Section IV. Finally, the conclusions and future works are summarized in Section V.

## II. PROBLEM FORMULATION

### A. System Architecture

The system architecture of the three-tier federated Cloud, Edge, and Fog computation offloading system is shown in Fig. 1, where UEs are considered as Fog servers. We consider a computation task is generated at a UE, and can be served by the UE itself, or offloads to the nearby edge server, or to the cloud server. We assume the cloud server is associated with $N$ edge servers and each edge server is associated with $M$ UEs. Tasks arrive at a UE according to a Poisson process. When a task arrives at a UE, with probability $p^{UE}$, it will be offloaded to the edge server. Similarly, when a task arrives at an edge server, with probability $p^{EC}$ it will be offloaded to the cloud server. This kind of offloading strategy is called probabilistic offloading which offload computation tasks to higher tier computing server based on a given probability.

When a task is offloaded to a higher-tier computing server, it is sent through a communication link (uplink). Depends on the size of the task and the bandwidth of the communication link, the task may encounter some communication delay. Similarly, after the computing server finishes the computation, the result of the task will be sent back through a downlink with some communication delay. Specifically, suppose the task is computed at the edge server. In that case, it will encounter an uplink communication delay from the UE to the edge server and a downlink communication delay in the opposite direction. Suppose the task is served at the cloud server. In that case, it will experience two uplink communication delays and two downlink communication delays, between the UE and the edge server and between the edge server and the cloud server, respectively.

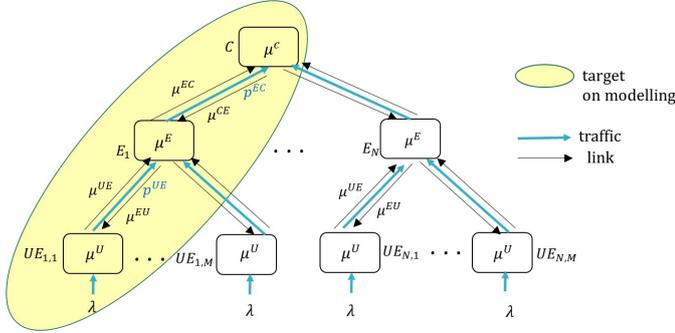

Fig. 1. The architecture of the three-tier federated Cloud, Edge, and Fog computation offloading system

### B. System Model

We analyze the delay of a computation task via queueing models. Specifically, we model each server and communication link as an M/M/1 queue and apply Jackson network theory to analyze a tandem of queues. Table I shows the notations used in our analysis. Fig. 2 shows the queueing network of the three-tier federated Cloud, Edge, and Fog computation system.

TABLE I
Table of Notations

| Notation | Meaning |
|---|---|
| $\mu^X$ ($\mu^C, \mu^E, \mu^U, \mu^{UE}, \mu^{EC}, \mu^{CE}, \mu^{EU}$) | Service capacity (rate) of the server or communication link X where X is either C (cloud), E (edge), U (UE), UE (uplink from UE to E), EC (uplink from edge to cloud), CE (downlink from cloud to edge), EU (downlink from edge to UE). |
| $\lambda^X$ ($\lambda^C, \lambda^E, \lambda^U, \lambda^{UE}, \lambda^{EC}, \lambda^{CE}, \lambda^{EU}$) | Arrival rate of the server or communication link X where X is either C (cloud), E (edge), U (UE), UE (uplink from UE to E), EC (uplink from edge to cloud), CE (downlink from cloud to edge), EU (downlink from edge to UE). |
| $\lambda$ | External task arrival rate to each UE. |
| $p^{UE}$ | Probability that a computation task will be offloaded from UE to the edge server. |
| $p^{EC}$ | Probability that a computation task will be offloaded from the edge server to the cloud server. |
| $v_X$ | The difference between service rate and arrival rate of X, i.e., $v_X = \mu^X - \lambda^X$, where X is either C, E, U, UE, EC, CE, or EU. The queuing delay of a M/M/1 server X is exponentially distributed with parameter $v_X$. |
| $\theta$ | Delay constraint of a computation task |
| $P_X^U(\theta)$ $P_X^E(\theta)$ $P_X^C(\theta)$ | The probability of the delay of a task served by UE, edge server, or cloud server exceeds the delay constraint θ respectively. |
| $P_X(\theta)$ | The overall probability of the delay of a task exceeds the delay constraint θ. |

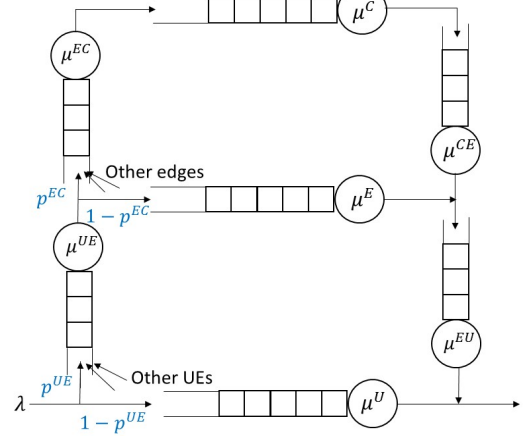

Fig. 2. Queueing network of the federated three-tier computation system.

In the following, we derive the probability that the delay experienced by a computation task is less than the delay constraint θ in three cases: namely, the task is either served by the UE, the edge server, or the cloud server.

*Case 1: the task is served by UE*

By assuming the arrival of computation tasks to a UE follows a Poisson process with rate $\lambda^U = (1 - p^{UE}) \times \lambda$, and the service time follows an exponential distribution with rate $\mu^U$, the delay of this M/M/1 queue at the UE follows an exponential distribution with rate $v^U = \mu^U - \lambda^U$. Thus, the probability of the delay of a task exceeds the delay constraint θ is given by

$$P_X^U(\theta) = P(X_U \geq \theta) = e^{-(\mu^U - \lambda^U)\theta}. \quad (1)$$

For ease of explanation, in this paper, we will assume a homogeneous scenario where all UEs have the same arrival and service rate. Similar assumptions apply to edge servers and each type of communication link. However, our results can be easily extended to a heterogeneous case.

*Case 2: the task is served by the edge server*

The delay consists of three parts: the delay of the uplink communication from UE to the edge server, the delay at the edge server, and the delay of the downlink communication from the edge server to the UE. Since an edge server is associated with $M$ UEs, the arrival rate of these three queues are the same, given by

$$\lambda^{UE} = \lambda^E = \lambda^{EU} = M \times (1 - p^{EC}) \times p^{UE} \times \lambda. \quad (2)$$

Since the output process of an M/M/1 queue is a Poisson process, the queues at the edge server and two communication links between UE and the edge server are all M/M/1 queues. Thus, the delay of a task served by an edge server can be derived by using Laplace transform. Let $X_E$ be the random variable of the delay. Let the three queues are indexed by $i$, $i=1, 2, or 3$. For example, queue 1 is the uplink from UE to the edge server, then $v_1 = \mu^{UE} - \lambda^{UE}$.

*Case 2-1: $v_i \neq v_j$ if $i \neq j$*

The Laplace transform of $X_E$ is given by

$$S^*(s) = \prod_{i=1}^{3} \frac{v_i}{s+v_i} = \prod_{i=1}^{3} v_i \times \prod_{i=1}^{3} \frac{1}{s+v_i}, \quad (3)$$

In general, if $X_E$ is the summation of $K$ exponential random variables with parameter $v_i, i = 1, ..., N$, then the Laplace transform of $X_E$ is given by

$$S^*(s) = \prod_{i=1}^{K} \frac{v_i}{s+v_i} = \prod_{i=1}^{K} v_i \times \prod_{i=1}^{K} \frac{1}{s+v_i}. \quad (4)$$

We show how to decompose equation (4) recursively and derive a general closed form for the inverse function (the delay distribution) of the Laplace transform. Let us start with $K=2$:

$$S^*(s) = \prod_{i=1}^{2} \frac{1}{s+v_i} = \frac{1}{v_2-v_1} \times \frac{1}{s+v_1} - \frac{1}{v_2-v_1} \times \frac{1}{s+v_2} \quad (5)$$

For $K=3$, $S^*(s)$ is given by

$$S^*(s) = \prod_{i=1}^{3} v_i \times \left[ \frac{1}{v_3-v_1} \times \prod_{i=1}^{2} \frac{1}{s+v_i} - \frac{1}{v_3-v_1} \times \prod_{i=2}^{3} \frac{1}{s+v_i} \right]$$

$$= \prod_{i=1}^{3} v_i \times \left[ \frac{1}{v_3-v_1} \times \frac{1}{v_2-v_1} \times \frac{1}{s+v_1} - \frac{1}{v_3-v_2} \times \frac{1}{v_2-v_1} \times \frac{1}{s+v_2} + \frac{1}{v_3-v_2} \times \frac{1}{v_3-v_1} \times \frac{1}{s+v_3} \right]. \quad (6)$$

And for any $K$, we can decompose $S^*(s)$ based on the result of $K-1$ by

$$S^*(s) = \prod_{i=1}^{K} v_i \times \left[ \begin{array}{l} \frac{1}{v_K-v_1} \times \prod_{i=1}^{K-1} \frac{1}{s+v_i} \\ -\frac{1}{v_K-v_1} \times \prod_{i=2}^{K} \frac{1}{s+v_i} \end{array} \right]. \quad (7)$$

For $K=3$, the delay distribution is obtained by inverting the Laplace transform, which is given by

$$f_X(x) = \prod_{i=1}^{3} v_i \times \left[ \frac{1}{v_3-v_1} \times \frac{1}{v_2-v_1} \times e^{-v_1 x} - \left( \frac{1}{v_3-v_2} \times \frac{1}{v_2-v_1} \right) \times e^{-v_2 x} + \frac{1}{v_3-v_1} \times \frac{1}{v_3-v_2} \times e^{-v_3 x} \right]. \quad (8)$$

Finally, the probability of the delay of a task exceeds the delay constraint θ is given by

$$P_X^E(\theta) = P(X_E \geq \theta) = \prod_{i=1}^{3} v_i \times \left[ \frac{1}{v_3-v_1} \times \frac{1}{v_2-v_1} \times \frac{1}{v_1} \times e^{-v_1\theta} - \left( \frac{1}{v_3-v_2} \times \frac{1}{v_2-v_1} \right) \times \frac{1}{v_2} \times e^{-v_2\theta} + \frac{1}{v_3-v_1} \times \frac{1}{v_3-v_2} \times \frac{1}{v_3} \times e^{-v_3\theta} \right]. \quad (9)$$

*Case 2-2: $v_i = v_j \; \forall i, j$*

The delay distribution becomes an Erlang distribution with parameter $r=3$. The probability of the delay of a task exceeds the delay constraint θ is given by

$$P_X^E(\theta) = P(X_E \geq \theta) = \sum_{i=0}^{2} \frac{(v_1\theta)^i}{i!} e^{-v_1\theta}. \quad (10)$$

*Case 2-3: $v_i = v_j \neq v_k$ (two $v_x$'s are the same while one is different)*

Without loss of generality, let $v_1 = v_2 \neq v_3$. Since two $v_i$'s are the same, the distribution of $X_E$ becomes the sum of an Erlang and an exponential distribution. Thus, the Laplace transform of $X_E$ is given by

$$S^*(s) = \prod_{i=1}^{3} v_i \times \frac{1}{(s+v_1)^2 \times (s+v_3)} = \frac{-1}{(v_3-v_1)^2} \times \frac{1}{s+v_1} + \frac{1}{(v_3-v_1)^2} \times \frac{1}{s+v_3} + \frac{1}{v_3-v_1} \times \frac{1}{(s+v_1)^2}. \quad (11)$$

And the probability of the delay of a task exceeds the delay constraint θ is given by

$$P_X^E(\theta) = P(X_E \geq \theta) = \prod_{i=1}^{3} v_i \times \left[ \begin{array}{l} \frac{1}{(v_3-v_1)^2} \times \frac{1}{v_3} \times e^{-v_3\theta} - \frac{1}{(v_3-v_1)^2} \times \frac{1}{v_1} \times e^{-v_1\theta} \\ + \frac{1}{v_3-v_1} \times \frac{1}{v_1^2} \times (e^{-v_1\theta} + v_1 x e^{-v_1\theta}) \end{array} \right]. \quad (12)$$

*Case 3: the task is served by the cloud server*

The delay consists of five parts: the delay of the uplink communication from UE to the edge server, the delay of the uplink communication from the edge server to the cloud server, the delay at the cloud server, the delay of the downlink communication from the cloud server to the edge server, and the delay of the downlink communication from the edge server to the UE. Since the cloud server is associated with $N$ edge servers, the arrival rates to the cloud server, the uplink from the edge server to the cloud server, and the downlink from the cloud server to the edge server are the same and given by

$$\lambda^{EC} = \lambda^C = \lambda^{CE} = N \times M \times p^{EC} \times p^{UE} \times \lambda, \quad (13)$$

while recall that $\lambda^{UE}, \lambda^{EU}$ are given in equation (2).

We have a tandem of five M/M/1 queues, and the delay of each queue is, again, an exponential distribution. Let us denote the parameter of the delay distribution by $v_i, i = 1, ..., 5$. For example, $v_1 = \mu^{UE} - \lambda^{UE}$ and $v_2 = \mu^{EC} - \lambda^{EC}$. And similar to case 2, depends on whether there are some $v_i$'s have the value, we have 7 sub-cases, denoted by (1,1,1,1,1), (5), (4,1), (3,1,1), (3,2), (2,1,1,1), (2,2,1), where each number denotes how many $v_i$'s are the same. For example, (3,2) denotes the case where three $v_i$'s have the same value while the other two $v_i$'s have the same value. Let us start from the case where each $v_i$ is different.

*Case 3-1: $v_i \neq v_j$ if $i \neq j$, denoted by (1,1,1,1,1)*

Based on equation (7), the Laplace transform of $X_C$ is given by

$$S^*(s) = \prod_{i=1}^{5} v_i \times \left[ \begin{array}{l} \frac{1}{v_5-v_1} \times \frac{1}{(s+v_1) \times (s+v_2) \times (s+v_3) \times (s+v_4)} \\ -\frac{1}{v_5-v_1} \times \frac{1}{(s+v_2) \times (s+v_3) \times (s+v_4) \times (s+v_5)} \end{array} \right]$$

$$= \prod_{i=1}^{5} v_i \times \left[ \begin{array}{l} \frac{1}{v_5-v_1} \times \frac{1}{v_4-v_1} \times \frac{1}{v_3-v_1} \times \frac{1}{v_2-v_1} \times \frac{1}{s+v_1} \\ -\frac{1}{v_5-v_2} \times \frac{1}{v_4-v_2} \times \frac{1}{v_3-v_2} \times \frac{1}{v_2-v_1} \times \frac{1}{s+v_2} \\ +\frac{1}{v_5-v_3} \times \frac{1}{v_4-v_3} \times \frac{1}{v_3-v_2} \times \frac{1}{v_3-v_1} \times \frac{1}{s+v_3} \\ -\frac{1}{v_5-v_4} \times \frac{1}{v_4-v_3} \times \frac{1}{v_4-v_2} \times \frac{1}{v_4-v_1} \times \frac{1}{s+v_4} \\ +\frac{1}{v_5-v_4} \times \frac{1}{v_5-v_3} \times \frac{1}{v_5-v_2} \times \frac{1}{v_5-v_1} \times \frac{1}{s+v_5} \end{array} \right]. \quad (12)$$

The delay distribution can be derived as in equation (8) and the probability of the delay of a task exceeds the delay constraint θ is given by

$$P_X^C(\theta) = P(X_C \geq \theta) = \prod_{i=1}^{5} v_i \times$$
$$\begin{bmatrix} \frac{1}{v_5-v_1} \times \frac{1}{v_4-v_1} \times \frac{1}{v_3-v_1} \times \frac{1}{v_2-v_1} \times \frac{1}{v_1} \times e^{-v_1\theta} \\ -\frac{1}{v_5-v_2} \times \frac{1}{v_4-v_2} \times \frac{1}{v_3-v_2} \times \frac{1}{v_2-v_1} \times \frac{1}{v_2} \times e^{-v_2\theta} \\ +\frac{1}{v_5-v_3} \times \frac{1}{v_4-v_3} \times \frac{1}{v_3-v_2} \times \frac{1}{v_3-v_1} \times \frac{1}{v_3} \times e^{-v_3\theta} \\ -\frac{1}{v_5-v_4} \times \frac{1}{v_4-v_3} \times \frac{1}{v_4-v_2} \times \frac{1}{v_4-v_1} \times \frac{1}{v_4} \times e^{-v_4\theta} \\ +\frac{1}{v_5-v_4} \times \frac{1}{v_5-v_3} \times \frac{1}{v_5-v_2} \times \frac{1}{v_5-v_1} \times \frac{1}{v_5} \times e^{-v_5\theta} \end{bmatrix}. \quad (13)$$

For the rest cases, the derivation is similar, so we only present the closed-form of the probability of the delay of a task exceeds the delay constraint θ.

*Case 3-2:* ($v_1 = v_2 = v_3 = v_4 = v_5$) denoted by (5)

For this case, the delay distribution follows an Erlang distribution with parameter *r=5*. The probability of the delay of a task exceeds the delay constraint θ is similar to equation (10) with the summation index *i* ranges from 0 to 4.

*Case 3-3:* ($v_1 = v_2 = v_3 = v_4 \neq v_5$) denoted by (4,1)

The probability $P_X^C(\theta)$ is given by
$$P_X^C(\theta) = P(X_C \geq \theta) = \prod_{i=1}^{5} v_i \times$$
$$\begin{bmatrix} \frac{-1}{(v_5-v_1)^4} \times \frac{1}{v_1} \times e^{-v_1\theta} \\ +\frac{1}{(v_5-v_1)^3} \times \frac{1}{v_1^2} \times (e^{-v_1\theta} + v_1\theta e^{-v_1\theta}) \\ -\frac{1}{(v_5-v_1)^2} \times \frac{1}{v_1^3} \times (e^{-v_1\theta} + v_1\theta e^{-v_1\theta} + \frac{(v_1\theta)^2}{2}e^{-v_1\theta}) \\ +\frac{1}{v_5-v_1} \times \frac{1}{v_1^4} \times (e^{-v_1\theta} + v_1\theta e^{-v_1\theta} + \frac{(v_1\theta)^2}{2}e^{-v_1\theta} \\ +\frac{(v_1\theta)^3}{6}e^{-v_1\theta}) + \frac{1}{(v_5-v_1)^4} \times \frac{1}{v_5} \times e^{-v_5\theta}) \end{bmatrix}. \quad (14)$$

*Case 3-4:* ($v_1 = v_2 = v_3 \neq v_4 \neq v_5$) denoted by (3,1,1)

The probability $P_X^C(\theta)$ is given by
$$P_X^C(\theta) = P(X_C \geq \theta) = \prod_{i=1}^{5} v_i \times$$
$$\begin{bmatrix} \frac{(v_4-v_1)^2+(v_5-v_1)^2+(v_4-v_1)(v_5-v_1)}{(v_4-v_1)^3(v_5-v_1)^3} \times \frac{1}{v_1} \times e^{-v_1\theta} \\ +\frac{2v_1-v_4-v_5}{(v_4-v_1)^2(v_5-v_1)^2} \times \frac{1}{v_1^2} \times (e^{-v_1\theta} + v_1\theta e^{-v_1\theta}) \\ +\frac{1}{(v_4-v_1)(v_5-v_1)} \times \frac{1}{v_1^3} \\ \times (e^{-v_1\theta} + v_1\theta e^{-v_1\theta} + \frac{(v_1\theta)^2}{2}e^{-v_1\theta}) \\ -\frac{1}{(v_4-v_1)^3(v_5-v_4)} \times \frac{1}{v_4} \times e^{-v_4\theta} \\ +\frac{1}{(v_5-v_1)^3(v_5-v_4)} \times \frac{1}{v_5} \times e^{-v_5\theta} \end{bmatrix}. \quad (15)$$

*Case 3-5:* ($v_1 = v_2 = v_3 \neq v_4 = v_5$) denoted by (3,2)

The probability $P_X^C(\theta)$ is given by
$$P_X^C(\theta) = P(X_C \geq \theta) = \prod_{i=1}^{5} v_i \times$$
$$\begin{bmatrix} \frac{3}{(v_4-v_1)^4} \times \frac{1}{v_1} \times e^{-v_1\theta} \\ -\frac{2}{(v_4-v_1)^3} \times \frac{1}{v_1^2} \times (e^{-v_1\theta} + v_1\theta e^{-v_1\theta}) \\ +\frac{1}{(v_4-v_1)^2} \times \frac{1}{v_1^3} \times \left(e^{-v_1\theta} + v_1\theta e^{-v_1\theta} + \frac{(v_1\theta)^2}{2}e^{-v_1\theta}\right) \\ -\frac{3}{(v_4-v_1)^4} \times \frac{1}{v_4} \times e^{-v_4\theta} \\ -\frac{1}{(v_4-v_1)^3} \times \frac{1}{v_4^2} \times (e^{-v_4\theta} + v_4\theta e^{-v_4\theta}) \end{bmatrix}. \quad (16)$$

*Case 3-6:* ($v_1 = v_2 \neq v_3 \neq v_4 \neq v_5$) denoted by (2,1,1,1)

The probability $P_X^C(\theta)$ is given by
$$P_X^C(\theta) = P(X_C \geq \theta) = \prod_{i=1}^{5} v_i \times$$
$$\begin{bmatrix} \left(\frac{-1}{(v_5-v_3)(v_4-v_3)(v_3-v_1)^2} \\ +\frac{1}{(v_5-v_4)(v_4-v_3)(v_4-v_1)^2} \\ -\frac{1}{(v_5-v_3)(v_5-v_4)(v_5-v_1)^2}\right) \times \frac{1}{v_1} \times e^{-v_1\theta} \\ +\left(\frac{1}{(v_5-v_3)(v_4-v_3)(v_3-v_1)} \\ -\frac{1}{(v_5-v_4)(v_4-v_3)(v_4-v_1)} \\ +\frac{1}{(v_5-v_3)(v_5-v_4)(v_5-v_1)}\right) \times \frac{1}{v_1^2} \times (e^{-v_1\theta} + v_1\theta e^{-v_1\theta}) \\ +\frac{1}{(v_5-v_3)(v_4-v_3)(v_3-v_1)^2} \times \frac{1}{v_3} \times e^{-v_3\theta} \\ -\frac{1}{(v_5-v_4)(v_4-v_3)(v_4-v_1)^2} \times \frac{1}{v_4} \times e^{-v_4\theta} \\ +\frac{1}{(v_5-v_3)(v_5-v_4)(v_5-v_1)^2} \times \frac{1}{v_5} \times e^{-v_5\theta} \end{bmatrix} \quad (17)$$

*Case 3-7:* ($v_1 = v_2 \neq v_3 = v_4 \neq v_5$) denoted by (2,2,1)

The probability $P_X^C(\theta)$ is given by
$$P_X^C(\theta) = P(X_C \geq \theta) = \prod_{i=1}^{5} v_i \times$$
$$\begin{bmatrix} \frac{3v_1-v_3-2v_5}{(v_3-v_1)^3(v_5-v_1)^2} \times \frac{1}{v_1} \times e^{-v_1\theta} \\ +\frac{1}{(v_3-v_1)^2(v_5-v_1)} \times \frac{1}{v_1^2} \times (e^{-v_1\theta} + v_1\theta e^{-v_1\theta}) \\ +\frac{v_1-3v_3+2v_5}{(v_3-v_1)^3(v_5-v_3)^2} \times \frac{1}{v_3} \times e^{-v_3\theta} \\ +\frac{1}{(v_3-v_1)^2(v_5-v_3)} \times \frac{1}{v_3^2} \times (e^{-v_3\theta} + v_3\theta e^{-v_3\theta}) \\ +\frac{1}{(v_5-v_1)^2(v_5-v_3)^2} \times \frac{1}{v_5} \times e^{-v_5\theta} \end{bmatrix}. \quad (18)$$

*Summary:*

Given the probability of the delay of a task exceeds the delay constraint θ in three cases, the overall probability of violating the delay constraint is given by
$$P_X(\theta) = P(X \geq \theta) = (1 - p^{UE}) \times P_X^U(\theta)$$
$$+(1 - p^{EC}) \times p^{UE} \times P_X^E(\theta) + p^{UE} \times p^{EC} \times P_X^C(\theta) \quad (19)$$

### C. Problem Statement

Given the following system parameters, how to configure $p^{UE}$ and $p^{EC}$ such that $P_X(\theta)$ could be minimized? The system parameters include number of UEs per edge server (*M*), number of edge servers per cloud server (*N*), external task arrival rate (*λ*), the service rate of each server and communication link ($\mu^C, \mu^E, \mu^U, \mu^{UE}, \mu^{EC}, \mu^{CE}, \mu^{EU}$), and the delay constraint θ.

### III. SUB-GRADIENT SEARCH (SGS) ALGORITHM

In this section, we proposed a search algorithm for finding the optimal $p^{UE}$ and $p^{EC}$ which minimizes $P_X(\theta)$ based on the sub-gradient method. Fig. 3 shows a typical relation among $p^{UE}$, $p^{EC}$ and $P_X(\theta)$, which clearly shows the convexity of $P_X(\theta)$. A formal proof is omitted in this paper due to the space limitation. But it is trivial to show that $P_X(\theta)$ us a convex function since the exponential function is convex, and $P_X(\theta)$ is either sum of some positive exponential functions or sum of both positive and negative exponential functions, but with more positive and larger exponential functions. This gives us the motivation to develop a sub-gradient search algorithm, referred to as SGS, as shown in Fig. 4. It adopts nonsummable diminishing step size. A sub-gradient method is an iterative method for solving a convex minimization problem.

The SGS algorithm, for a given $p^{UE}$, first searches for the optimal $p^{EC}$. After the $p^{EC}$ is set, it then searches for the optimal $p^{UE}$. It then repeats until both $p^{UE}$ and $p^{EC}$ converge to a fixed point. The main idea of the search algorithm is to set the search direction according to the direction that leads to lower $P_X(\theta)$ which follows the sub-gradient method for solving convex optimization.

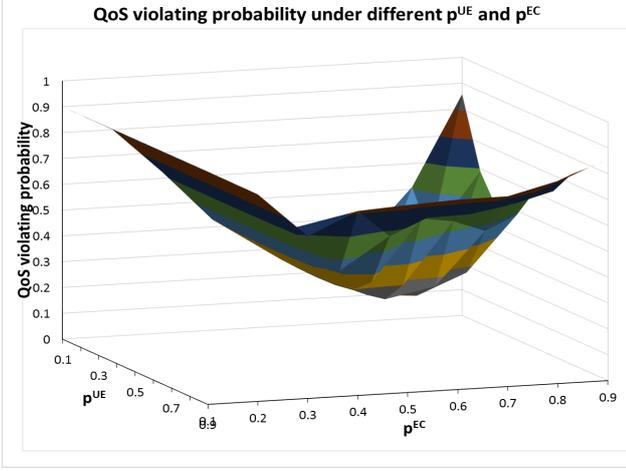

Fig. 3. $P_X(\theta)$ as a function of $p^{UE}$ and $p^{EC}$.

| Sub-Gradient Search Algorithm |
| --- |
| **Input:** M, N, $\lambda$, $\mu^C, \mu^E, \mu^U, \mu^{UE}, \mu^{EC}, \mu^{CE}, \mu^{EU}, \theta$ <br> **Output:** optimal $p^{UE}, p^{EC}$ <br> **Begin** <br>    $p^{UE} = 0.5; p^{EC} = 0.5;$ //initialize $p^{UE}, p^{EC}$ <br>    Repeat { <br>       $old\_p^{UE} = p^{UE}; old\_p^{EC} = p^{EC};$ <br>       step=0.25; k=1; //iteration number <br>       $new\_p^{EC} = p^{EC} + step;$ <br>       Repeat { //search for optimal $p^{EC}$ <br>          $old\_P_X(\theta)$ = calculate $P_X(\theta)$ using equation (19) and $p^{UE}, p^{EC};$ <br>          $new\_P_X(\theta)$ = calculate $P_X(\theta)$ using equation (19) and $p^{UE}, new\_p^{EC};$ <br>          If ($\|new\_P_X(\theta) - old\_P_X(\theta)\|$<0.000001) <br>            break; // exit repeat loop <br>          else if $old\_P_X(\theta) < new\_P_X(\theta)$ { <br>            //change $new\_p^{EC}$ to get closer to $p^{EC}$ <br>            if ($p^{EC} < new\_p^{EC}$) { <br>               $new\_p^{EC} = p^{EC} - step;$if $new\_p^{EC} < 0$ $new\_p^{EC} = 0;$ <br>            } else { <br>               $new\_p^{EC} = p^{EC} + step;$if $new\_p^{EC} > 1$ $new\_p^{EC} = 1;$ <br>            } <br>          } else { //set $p^{EC}$ to $new\_p^{EC}$, try a new search from new $p^{EC}$ <br>            $p^{EC}\_old = p^{EC}; p^{EC} = new\_p^{EC};$ <br>            if ($new\_p^{EC} < p^{EC}\_old$) { <br>               $new\_p^{EC} = p^{EC} - step;$if $new\_p^{EC} < 0$ $new\_p^{EC} = 0;$ <br>            } else { <br>               $new\_p^{EC} = p^{EC} + step;$if $new\_p^{EC} > 1$ $new\_p^{EC} = 1;$ <br>            } <br>          } <br>          k++; step = $0.25/\sqrt{k}$; // Nonsummable diminishing step size <br>       } until (step<0.0001); //until step is less than threshold <br>       step=0.25; k=1; <br>       $new\_p^{UE} = p^{UE} + step;$ <br>       Repeat { //search for optimal $p^{UE}$ <br>          Update $new\_p^{UE}$ and $p^{UE}$ as the update procedure for $new\_p^{EC}$ and $p^{EC}$ in the above repeat loop; <br>       } until (step<0.0001); <br>    } until (($\|old\_p^{UE} - p^{UE}\|$<0.0001) && ($\|old\_p^{EC} - p^{EC}\|$<0.0001)); <br> **End** |

Fig. 4. Pseudocode for the SGS algorithm.

## IV. NUMERICAL RESULTS

### A. System Parameters

The setting of system parameters is shown in Table II, which is used to validate our analytical results as well as the optimality of the SGS algorithm. The time unit shall be set according to the real-world scenario. Notably, without offloading to higher tier servers, the arrival rate exceeds the service capacity of UEs. Offloading all traffic to the edge or cloud server also exceeds their service capacity.

TABLE II
System parameters (rates are per time unit) (M=N=5, $\theta = 1.2$)

| $\lambda$=2 | $\mu^C$=25 | $\mu^E$=8 | $\mu^U$=1.5 |
| --- | --- | --- | --- |
| $\mu^{UE}$=12 | $\mu^{EC}$=22 | $\mu^{CE}$=21 | $\mu^{EU}$=11 |

We first validate our analytical results by comparing them with the simulation results, as shown in Fig. 5 where $p^{EC}$ is set to 0.4 and $p^{UE}$ ranges from 0.1 to 0.9. Each simulation is run for 30 runs with a simulation time of 10000 time units, and the 95% confidence interval is less than 0.5% of the mean value.

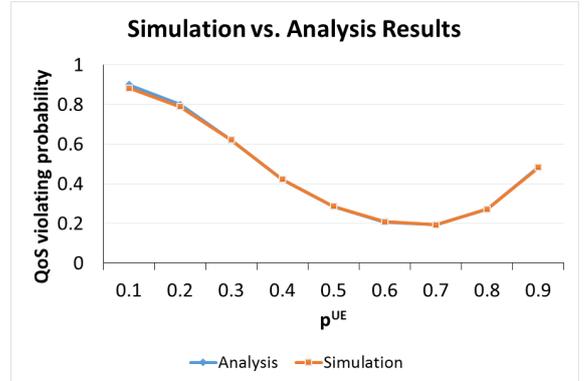

Fig. 5. Comparison of simulation results with analytical results.

Fig. 6 shows the analytical result of QoS violating probability at each tier with the same setting in Fig. 5. As $p^{UE}$ increases, more tasks are offloaded to the higher tier server. As a consequence, the QoS violating probability at edge and cloud server increases as $p^{EC}$ increases while that of UE decreases. At the lowest QoS violating probability case, the QoS violating probabilities at UE, edge, and cloud are 0.340, 0.121, 0.146, respectively. In addition, our simulation results showed that the average end-to-end delays for tasks serving by UE, edge, and cloud are 1.11, 0.71, 0.81, respectively, all less than $\theta$ (1.2). This shows the importance of delay distribution analysis as even with more than 19% of tasks cannot meet their deadline, their mean delays are still less than the delay constraint.

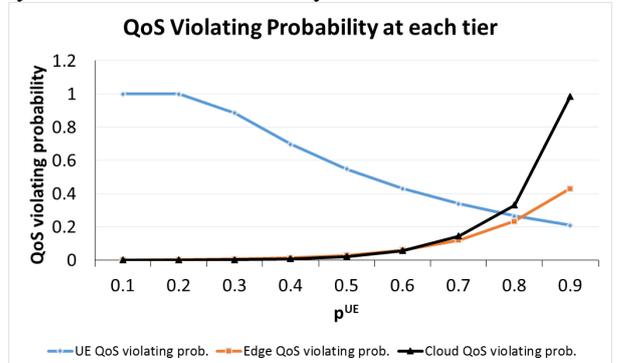

Fig. 6. Comparison of QoS violating probability at different tier

Secondly, we validate the SGS algorithm. System parameters are set as in Table II, and we let $p^{UE}$ and $p^{EC}$ varies from 0.1 to 0.9 with an interval of 0.1. The QoS violating probability vs. $p^{UE}$ and $p^{EC}$ is plotted in Fig. 3 and the lowest QoS violating probability is 0.194, which occurs when $p^{UE} = 0.7$ and $p^{EC} = 0.4$. Our SGS algorithm yields the optimal QoS violating probability of 0.188 when $p^{UE} = 0.675$ and $p^{EC} = 0.370$. The result validates that the SGS algorithm is able to find the optimal $p^{UE}$ and $p^{EC}$. We also run a simulation with $p^{UE} = 0.675$ and $p^{EC} = 0.370$ and the result of QoS violating probability is 0.189, which matches the result of the SGS algorithm well.

Fig. 7 shows how the external arrival rate affects the QoS violating probability and the optimal offloading probabilities. Intuitively, as the arrival rate increases, so is the QoS violating probability. Interestingly, as the arrival rate increases, fewer tasks are offloaded to higher tier servers. Finally, Fig. 8 shows when the service capacity of the edge server changes, how it affects the optimal configuration of $p^{UE}$ and $p^{EC}$, as well as the QoS violating probability. As we can see, as the edge server capacity increases, more tasks are offloaded to the edge server, so the $p^{UE}$ increases while $p^{EC}$ decreases. And since the system has more capacity, the overall QoS violating probability also decreases.

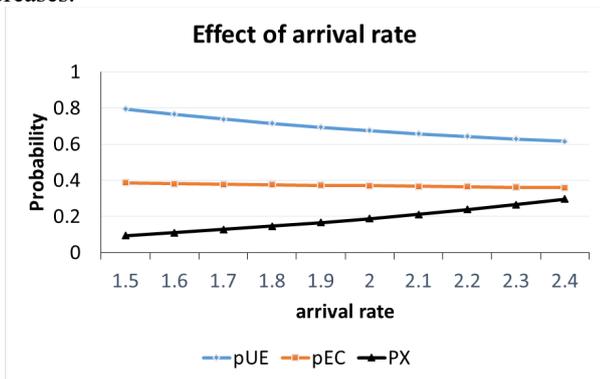

Fig. 7. The effect of external arrival rate on QoS violating probability

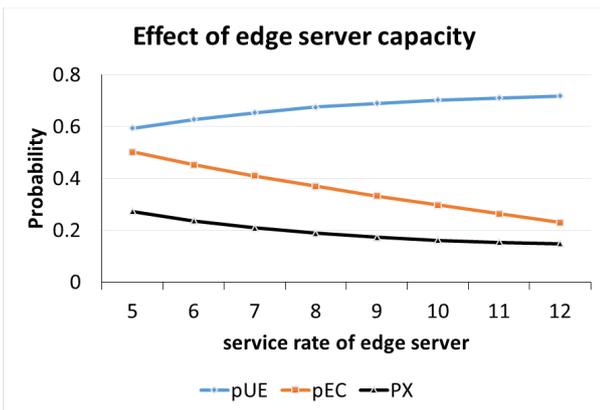

Fig. 8. The effect of edge server capacity on $p^{UE}$, $p^{EC}$ and QoS violating probability

## V. CONCLUSIONS

This paper investigated the delay constraint problem in a three-tier computation offloading system with the cloud server, edge server, and fog server (mobile device). Closed-form solutions have been derived for the delay distribution, and the probability of the delay of a task exceeds the delay constraint θ. For the probabilistic offloading strategy, we proposed a sub-gradient-based search (SGS) algorithm to find the optimal offloading probabilities while minimizing the delay violating probability. Our numerical results showed that our modeling is valid, and the SGS algorithm is able to find the optimal offloading probabilities.

Several works need further investigation. First, more experiments are needed to explore the effect of several system parameters. Second, we are investigating an online queue-length-based offloading strategy that offloads a task to a higher tier computation server when it arrives at a lower-tier server and finds the probability that the delay of serving by this lower tier server will exceed the delay constraint is larger than some threshold.